# Assessing the Validity of a *a priori* Patient-Trial Generalizability Score using Real-world Data from a Large Clinical Data Research Network: A Colorectal Cancer Clinical Trial Case Study

**Qian Li, MS[1#], Zhe He, PhD[2#], Yi Guo, PhD[1#], Hansi Zhang, MS[1], Thomas J George Jr, MD, FACP[1], William Hogan, MD, MS[1], Neil Charness, PhD[2], Jiang Bian, PhD[1\*]**
[1]University of Florida, Gainesville, FL, USA; [2]Florida State University, Tallahassee, FL, USA;

**Abstract**

*Existing trials had not taken enough consideration of their population representativeness, which can lower the effectiveness when the treatment is applied in real-world clinical practice. We analyzed the eligibility criteria of Bevacizumab colorectal cancer treatment trials, assessed their a priori generalizability, and examined how it affects patient outcomes when applied in real-world clinical settings. To do so, we extracted patient-level data from a large collection of electronic health records (EHRs) from the OneFlorida consortium. We built a zero-inflated negative binomial model using a composite patient-trial generalizability (cPTG) score to predict patients' clinical outcomes (i.e., number of serious adverse events, [SAEs]). Our study results provide a body of evidence that 1) the cPTG scores can predict patient outcomes; and 2) patients who are more similar to the study population in the trials that were used to develop the treatment will have a significantly lower possibility to experience serious adverse events.*

**Introduction**

Clinical studies (trials) are essential in evidence-based medicine.[1] Clinical trials, however, are often conducted under idealized and rigorously controlled conditions to improve their internal validity and success rates; but such conditions, paradoxically, may compromise their external validity (i.e., trial results' generalizability to the real-world target populations).[2] These idealized conditions are sometimes exaggerated and reflected as overly restrictive eligibility criteria.[2] The generalizability and study population representativeness have long been major concerns.[2,3] Certain population subgroups, such as older adults,[4,5] are often underrepresented due to unjustified exclusion criteria, especially in cancer studies[6–10]. The underrepresentation of these population subgroups can lead to low trial generalizability, and subsequently, reduce treatment effectiveness and increase the likelihood of adverse outcomes in these population subgroups when the treatments are moved into real-world clinical practice. As a consequence, some approved drugs have been withdrawn from the market after serious adverse drug reactions were observed on population subgroups excluded from the original trials.[11]

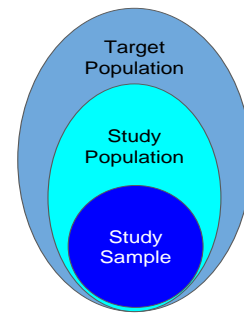

In clinical trials, the target population represents the patients to whom the results of the clinical trials are intended to be applied. The study population represents the patients being sought as defined in the clinical trial eligibility criteria (***Figure 1***). To ensure patient safety and demonstrate treatment efficacy, eligibility criteria are often restrictive, thus representing a constrained subset of the target population. Further, study samples are the enrolled participants of a trial. Even though study participants are screened based on eligibility criteria, due to real-world constraints such as trial awareness, the study sample may not adequately represent the study population defined by the eligibility criteria. Further, it is worth noting that the notions of generalizability and population representativeness are related but distinct. Population representativeness measures the study population's coverage of real-world patients, with respect to study traits (e.g., age, vitals, and labs), often defined by eligibility criteria. On the other hand, generalizability is the portability of the causal effects of an intervention to real-world settings. Besides population representativeness, other factors also affect studies' generalizability, such as variation in patients across different clinical settings, discrepancies in conditions under which a trial was conducted, and incomplete reporting.[12,13] Nevertheless, population representativeness of the study population is one of the determining factors for its generalizability.

**Figure 1**. Populations in clinical *trials.*

---

[#] Contributed equally, co-first authors
[*] Corresponding: Jiang Bian; bianjiang@ufl.edu

There are two major types of approaches to assessing a study's population representativeness: 1) the *a priori* generalizability is the representativeness of eligible participants (study population) to the target population; 2) the *a posteriori* generalizability is the representativeness of enrolled participants (study sample) to the target population. The *a priori* generalizability can also be called as the eligibility-driven generalizability, whereas the *a posteriori* generalizability can be called as the sample-driven generalizability.[14] However, the *a posteriori* generalizability is an artifact after the fact (i.e., can only be done when a trial is concluded and complete trial data are collected), while the *a priori* generalizability is affected by the trial's eligibility criteria that are modifiable during the study design phase.

In 2014, an *a priori* generalizability score "Generalizability Index for Study Traits" (GIST 1.0; we use 1.0 to differentiate these from the GIST 2.0 metrics introduced below) was introduced to quantify the population representativeness using eligibility criteria one at a time and real-world patient data.[15] The GIST 1.0 score characterizes the proportion of patients that would be potentially eligible across trials with the same trait over the target population. GIST 1.0 was validated using a simulated target population.[16] As some criteria are correlated, GIST was extended to mGIST 1.0 with joint use of multiple eligibility criteria.[17] Later, GIST 2.0[18] was developed to consider both 1) the dependencies across multiple criteria in a study, and 2) the significance of individual traits across different diseases (e.g., HbA1C is more important in type 2 diabetes than it is in chronic kidney disease). GIST 2.0 has two components: sGIST (w.r.t. one criterion) and mGIST (w.r.t. multiple criteria) corresponding to the original GIST 1.0 and mGIST 1.0, respectively. GIST scores are between 0 and 1, where a higher score indicates a greater population representativeness. From now on, we will use sGIST and mGIST to refer to the GIST 2.0 metrics.

Nevertheless, to be able to rationalize adjustments of eligibility criteria towards ultimately better generalizability early on, it is important to identify the relationships among the *a priori* generalizability and the actual outcomes of the interventions/treatments in real-world settings. Treatment outcomes can be measured in several different ways: 1) improvement in clinical outcomes (e.g., better lab results), 2) less adverse events (AEs); and ultimately 3) longer survival and better quality-of-life (QoL). Adverse event is an important measure of treatment safety during clinical trials. Previously, Sen et al. found that GIST 2.0 score was significantly correlated with the number of AEs (i.e., the lower the GIST 2.0 score the higher the number of AEs) based on data from 16 sepsis trial results sections in ClinicalTrials.gov.[14] Nevertheless, their study is limited as it does not provide any evidence on whether the *a priori* generalizability of the trials had any impact on the clinical outcomes (i.e., the number of AEs) when the treatment is applied to the target patients in real-world clinical practices.

On the other hand, the wide adoption of electronic health record (EHR) systems and the proliferation of clinical data warehouses with rich real-world patient datasets offer unique opportunities to address these studies. The U.S. Food and Drug Administration (FDA) recently coined the terms real-world evidence (RWE) and real-world data (RWD) as "*data regarding the usage, or the potential benefits or risks, of a drug derived from sources other than traditional clinical trials.*"[19,20] RWD can come from various sources including EHRs, claims and billing activities as well as patient-generated data. These data will play an increasingly important role in health care and regulatory decisions.

In this study, we aim to fill this important knowledge gap exploring the hypotheses that 1) the *a priori* generalizability of treatment trials is correlated with the clinical outcomes of the treatment (i.e., the number of AEs) and 2) patients who are eligible (i.e., based on the trials' eligibility criteria and patient characteristics in their EHRs) for the original trials used to develop the treatment will have better clinical outcomes than those who are not, when the treatment is applied on both eligible/ineligible patients in real-world clinical settings, using RWD—a large collection of linked EHRs and claims—from a large clinical data research network. Our current study focuses on Bevacizumab (under the trade name Avastin)—a first of its kind of monoclonal antibody as a tumor-starving (anti-angiogenic) therapy—approved by the FDA in 2006 for the treatment of metastatic colorectal cancer.[21]

**Methods**

*Data sources*

***Bevacizumab (Avastin) clinical trials and trial eligibility criteria***. We obtained free-text eligibility criteria from ClinicalTrials.gov—a registry maintained by the National Library of Medicine (NLM) in the United States. As of March 2019, over 299,335 studies across all 50 states in the US as well as in 208 countries are registered on ClinicalTrials.gov. Study information in ClinicalTrials.gov is semi-structured: study descriptors such as study phase, intervention type, and locations are stored in structured fields while eligibility criteria are largely free-text. Through ClinicalTrials.gov, we found 57 Bevacizumab trials that met our inclusion criteria: 1) the trial was conducted in the US; 2) the primary purpose of the trial is the development of the treatment agent; and 3) excluding post-market observational studies (e.g., comparative effectiveness studies, and essentially all Phase IV studies).

*Real-world patient data from the OneFlorida.* We obtained individual-level patient data from the OneFlorida Clinical Research Consortium (OneFlorida CRC),[22] one of the 13 Clinical Data Research Networks (CDRNs) contributing to the national Patient-Centered Clinical Research Network (PCORnet) previously funded by the Patient-Centered Outcomes Research Institute (PCORI). PCORnet is now supported by the newly incorporated People Centered Research Foundation (PCRF), a nonprofit formed by PCORnet investigators with significant additional infrastructure-building funds from PCORI. The OneFlorida data repository integrated various data sources from contributing organizations in the OneFlorida CRC currently including 10 healthcare organizations: 1) two academic health centers (i.e., University of Florida Health, UFHealth and University of Miami Health System, UHealth), 2) seven healthcare systems including Tallahassee Memorial Healthcare (TMH affiliated with Florida State University), Orlando Health (ORH), Adventist Health (AH, formerly known as Florida Hospital), Nicklaus Children's Hospital (NCH, formerly known as Miami Children's Hospital), Bond Community Health (BCH), Capital Health Plan (CHP), and Health Choice Network, (HCN), and 3) CommunityHealth IT—a rural health network in Florida. In addition, OneFlorida also obtained claims data from the Florida Medicaid (FLM) program. As a network, the OneFlorida CRC provides care for approximately 48% of Floridians through 4,100 physicians, 914 clinical practices, and 22 hospitals with a catchment area covering all 67 Florida counties.[23] Most HCOs in OneFlorida contributed EHRs, while CHP and FLM contributed claims data. We linked patients across the different EHRs and claims data sources using a validated privacy-preserving record linkage method.[24,25] OneFlorida contains only a limited data set under the Health Insurance Portability and Accountability Act (HIPAA) and follows the PCORnet Common Data Model (CDM) v4.1 including patient demographics, enrollment status, vital signs, conditions, encounters, diagnoses, procedures, prescribing (i.e., provider orders for medications), dispensing (i.e., outpatient pharmacy dispensing), and lab results. The scale of the data is ever growing with a collection of longitudinal and robust patient-level records of ~15 million Floridians and over 463 million encounters, 917.6 million diagnoses, 1 billion prescribing records, and 1.17 billion procedures as of December 2018. Since our goal is to evaluate the generalizability of Bevacizumab trials as a colorectal cancer treatment, we extracted colorectal patients from OneFlorida as the target population using ICD-9/10-CM codes (i.e., ICD-9: 153.\*, 154.\*, 159.0; ICD-10: C18.\*, C19.\*, C20.\*, C26.0). We identified 39,776 unique colorectal patients and extracted their data from OneFlorida.

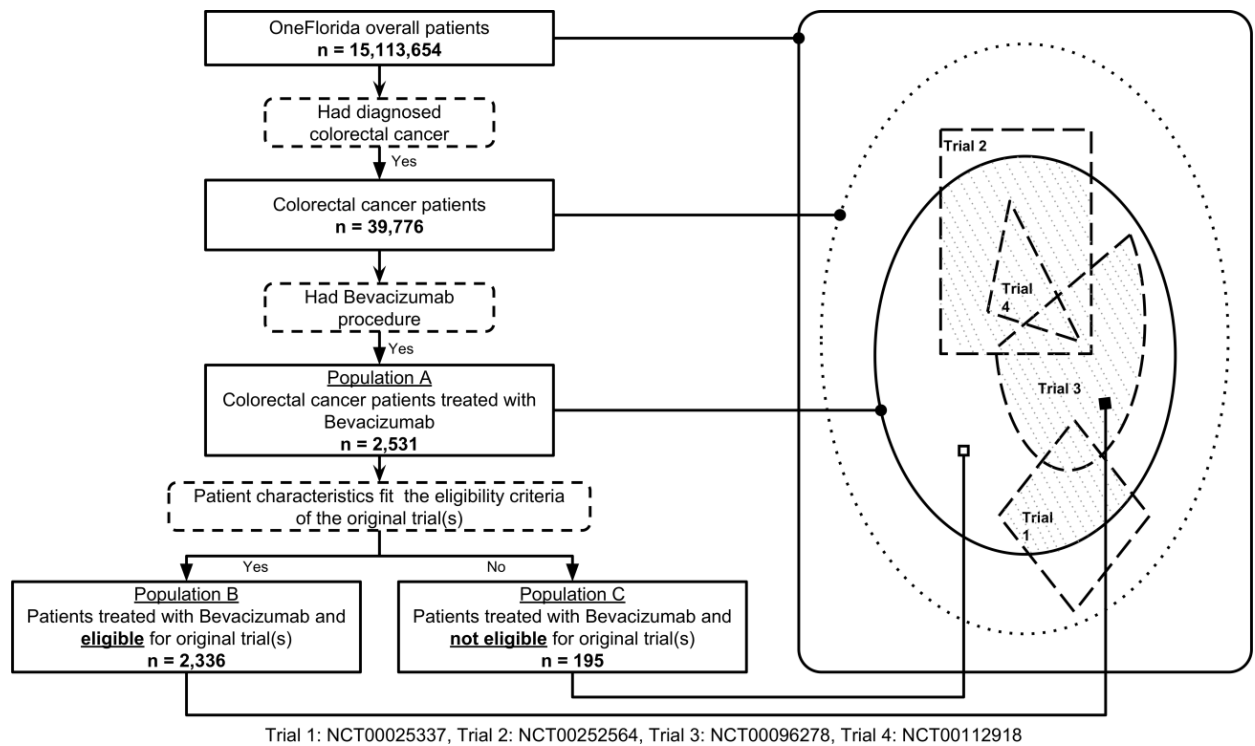

**Figure 2**. Overall study design and selection of study populations.

*Overall study design*

There are 3 different populations in our study: 1) <u>population A</u>: OneFlorida colorectal cancer patients who were treated with Bevacizumab; 2) <u>population B</u>: patients who were treated with Bevacizumab and eligible for the original trials that were used to develop the treatment agent (i.e., a subset of population A based on the eligibility criteria of the original trials and patient characteristics defined by their EHRs in OneFlorida); and 3) population C: patients who were treated with Bevacizumab but NOT eligible for the original trials (i.e., a subset of A). The selection of and relationships among these populations are illustrated in *Figure 2*.

We aimed to compare the clinical outcomes (i.e., the number AEs) of population B and population C considering both 1) the *a priori* generalizability of the original Bevacizumab trials and 2) whether the patients can be eligible for the original trials based on their EHRs. Our hypothesis is that population B will have better outcomes (i.e., less AEs) compared with population C. To do so, we devised a composite patient-trial generalizability (cPTG) score based on GIST that considers both trial generalizability and patient eligibility of the individual trials.

Our analysis consists of 7 steps: 1) analyzing Bevacizumab colorectal cancer treatment trials extracted from ClinicalTrials.gov to determine the computability of each eligibility criterion and constructing queries to extract study traits corresponding to each trial eligibility criterion based on the OneFlorida data; 2) identifying the different populations of interest (i.e., populations B and C as described above); 3) calculating each trial's GIST score based on the computable eligibility criteria of each trial; 4) determining the eligibility (i.e., a binary variable) of each patient for each Bevacizumab trial of interest; 5) calculating the composite patient-trial generalizability score for each patient; 6) identifying each patient's clinical outcome (i.e., the number of serious AEs) of being treated with Bevacizumab; and 7) comparing the difference in the numbers of serious AEs between population B and population C and examining the relationships between the cPTG score and the number of serious AEs.

**Table 1.** ICD-9/10 codes to define serious adverse events for Bevacizumab based on its drug label.

| Serious Adverse Event | ICD-9-CM | ICD-10-CM |
|---|---|---|
| Asymptomatic postprocedural ovarian failure | 256.2 | E89.40 |
| Systolic (congestive) heart failure | 428.2 | I50.2 |
| Diastolic (congestive) heart failure | 428.3 | I50.3 |
| Combined systolic and diastolic heart failure | 428.4 | I50.4 |
| Cerebral Hemorrhage | 431 | I60 |
| Intracranial Hemorrhage | 432.9 | I62.9 |
| Fistula of intestine | 569.81 | K63.2 |
| Perforation of intestine | 569.83 | K63.1 |
| Gastrointestinal hemorrhage | 578 | K92.2 |
| Hematemesis | 578.0 | K92.0 |
| Vaginal bleeding | 626.8 | N93.9 |
| Epistaxis | 784.7 | R04.0 |
| Hemoptysis | 786.3 | R04.2 |
| Proteinuria | 791.0 | R80 |

*Defining populations of interest and serious adverse events (SAEs) related to using Bevacizumab*

We used International Classification of Diseases, Ninth Revision/Tenth Revision, Clinical Modification (ICD-9/10-CM) codes to identify patients who were diagnosed with colorectal cancer in the OneFlorida data. Within the colorectal cancer patients, we then used the Healthcare Common Procedure Coding System (HCPCS) codes (e.g., C9257, J9035) combined with Bevacizumab's RxNORM (e.g., 337521) codes to identify the administration of Bevacizumab. To identify SAEs in patients treated with Bevacizumab, we first reviewed the FDA approved drug label of Bevacizumab using the DailyMed database maintained by NLM. Then for each SAE, we identify the corresponding ICD-9/10-CM codes, as shown in **Table 1**. As shown in **Figure 3**, to count as a SAE related to Bevacizumab, the SAE diagnosis code has to occur 1) after the first Bevacizumab procedure, but 2) within 180 days after the last Bevacizumab procedure. Note that the first Bevacizumab procedure has to occur after the diagnosis of colorectal cancer. Using these criteria, we counted the total number of SAEs for each patient.

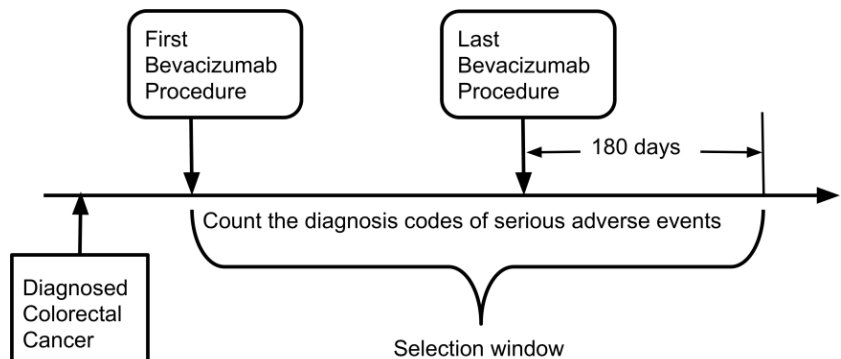

**Figure 3**. Selection window for serious adverse events related to treating colorectal cancer with Bevacizumab.

*A new composite patient-trial generalizability (cPTG) score based on GIST*

The original GIST metric quantifies the *a priori* generalizability of clinical trials with respect to selected quantitative eligibility criteria that specify a permissible value range (e.g., HbA1c > 7%), one at a time.[15] The GIST score ranges from 0 to 1, with 0 being not generalizable and 1 being perfectly generalizable. In essence, it characterizes the proportion of patients potentially eligible across trials. The extension of GIST, mGIST,[17] can quantify the population representativeness with joint use of multiple criteria of interest. Both GIST and mGIST focus on the generalizability assessment at the disease domain level (i.e., assessing the generalizability of trials targeting the same disease). GIST 2.0 was then introduced as a scalable multivariate metric for quantifying the population representativeness of individual clinical trials by explicitly modeling the dependencies among all eligibility criteria.[18] The original implementation of GIST 2.0 was in Matlab. We implemented GIST 2.0 in Python and made it available in a public Github repository (i.e., https://github.com/Andeeli/GIST.git). The calculation of sGIST and mGIST is trivial and details can be found in the original GIST 2.0 publication.[18] One key input is to define the target population (i.e., the real-world patient population that the treatment is intended to be applied on). For our study, we defined patients with colorectal cancer and treated with Bevacizumab (i.e., population A) as our target population. The outputs of GIST 2.0 are single-trait GIST score (sGIST) for each trait and one multiple-trait GIST score (mGIST) for the trial.

As the mGIST score is a trial-level variable, we also need to define whether a patient is thought to be eligible of a trial based on its eligibility criteria and corresponding patient traits as defined in their EHRs (i.e., based on OneFlorida data in our case). The process is straightforward. For example, "*Platelet count at least 1,500/mm^3*" is an inclusion criterion in trial NCT00025337. The lab results for platelet tests are coded using Logical Observation Identifiers Names and Codes (LOINC) (i.e., 26515-7, 777-3, and 778-1) in OneFlorida. Based on these LOINC codes, we queried patients' platelet test results and transformed the measurement unit to mm/^3 if needed. If a patient's platelet lab result is larger or equal to 1,500/mm^3, we will consider the patient as met this particular inclusion criterion. We then determined the patient's eligibility (i.e., a binary variable) of the trial, when the patient met all the inclusion criteria while did not meet any of the exclusion criteria. Note that not all eligibility criteria are computable as the needed data elements may not exist in their EHRs (e.g., "*Fertile patients must use effective contraception*"). We did not use these non-computable criteria when determining a patient's eligibility.

In order to consider both the patient-level eligibility and trial-level *a priori* generalizability, we propose a composite patient-trial generalizability (cPTG) score. To calculate cPTG score, we first calculated the mGIST score (ranging from 0 to 1) for each trial. Then, we used patients' traits data (from OneFlorida) to create an index (i.e., 0 as not eligible or 1 as eligible) to indicate whether a patient is eligible for the specific trial or not. We then took the average of the dot product of the vector of mGIST and the vector of patient's eligibility as follows:

$$cPTG_i = \frac{1}{K}\sum_{j}^{K} e_{ij} g_j$$

where $e_{ij}$ is the eligibility (0 or 1) for patient *i* of trial *j*, and $g_j$ is the generalizability score (i.e., mGIST) of trial *j*, for each patient *i* (i.e., *i=1,2,…,N*) and each trial *j* (i.e., *j=1,2,…,K*). The cPTG score ranges from 0 to 1. Intuitively, a higher cPTG score for a patient means the patient is eligible for more trials in the set of trials of interest and those trials have higher *a priori* generalizability.

*Statistical analysis*

Mean and standard deviation were calculated for continuous variables. For categorical variables, frequency and percentage was calculated. As more than 80% of the patients in our data have zero adverse events. We fit a zero-inflated model to consider the number of SAEs as the outcome and the cPTG score as a predictor. We also considered the following variables as controlling covariates in the model: the number of Bevacizumab procedure, days of follow up from latest Bevacizumab procedure, and patient demographics including age, gender, race, and ethnicity. Since the variance of the outcome is much larger than the mean, the data is over-dispersed. So, we preferred the zero-inflated negative binomial model to a zero-inflated Poisson regression model.[26]

**Results**

**Cohort characteristics**

We identified 2,531 unique patients who had been diagnosed with colorectal cancer and treated with Bevacizumab (out of a total of 39,776 colorectal cancer patients). Among these patients, 2,034 (80.4%) had no SAE while 497 (19.6%) had at least one SAE. The average age of the patient at her/his last Bevacizumab procedure is 59 years old. Although male patients are slightly more than female patients, there are more female patients with SAEs. **Table 2** shows the characteristics of our target population.

Table 2. Demographic characteristics and outcomes of the target population in OneFlorida.

|  | Overall (N=2,531) | | # of SAEs = 0 (N=2,034) | | # of SAE > 0 (N=497) | |
|---|---|---|---|---|---|---|
|  | N (or Mean) | % (or SD) | N (or Mean) | % (or SD) | N (or Mean) | % (or SD) |
| **Age at last PX (years)** | 59.09 | 11.4 | 59.46 | 11.4 | 57.59 | 11.29 |
| **Gender** | | | | | | |
| Female | 1,206 | 47.6% | 935 | 46.0% | 271 | 54.5% |
| Male | 1,325 | 52.4% | 1099 | 54.0% | 226 | 45.5% |
| **Race/Ethnicity** | | | | | | |
| Non-Hispanic White | 1,069 | 42.2% | 868 | 42.7% | 201 | 40.4% |
| Non-Hispanic Black | 481 | 19.0% | 377 | 18.5% | 104 | 20.9% |
| Hispanic | 528 | 20.9% | 407 | 20.0% | 121 | 24.3% |
| Other | 17 | 0.7% | 15 | 0.7% | 2 | 0.4% |
| Unknown | 436 | 17.2% | 367 | 18.0% | 69 | 13.9% |
| **cPTG** | 0.49 | 0.20 | 0.50 | 0.20 | 0.45 | 0.21 |
| **Number of SAEs** | 0.98 | 4.08 | 0 | 0 | 4.99 | 8.04 |
| **Follow up days** | 103.1 | 66.97 | 99.95 | 67.91 | 115.96 | 61.4 |
| **Number of Bevacizumab PXs** | 10.78 | 11.63 | 9.96 | 10.86 | 14.17 | 13.87 |
| **First PX to Last PX in days** | 263.52 | 328.17 | 234.75 | 296.8 | 381.21 | 413.49 |

*PX: Procedure; SAE: Serious Adverse Event; SD: Standard Deviation; cPTG: composite patient-trial generalizability

**Analysis of Bevacizumab colorectal cancer trial eligibility criteria**

From the collection of 57 Bevacizumab colorectal trials, we extracted 1,674 eligibility criteria (i.e., 951 inclusion criteria and 723 exclusion criteria) from ClinicalTrails.gov. 124 of the 951 (13.04%) inclusion criteria contained negations; and 19 of the 723 (2.63%) exclusion criteria contained negations. On average, each colorectal trial has 26 (4 to 65) inclusion criteria and 27 (0 to 39) exclusion criteria. Comparing to our previous study on Hepatitis C Virus (HCV) trials[27], colorectal cancer trials have significantly more inclusion and exclusion criteria (5.56 inclusion criteria and 7.98 exclusion criteria in HCV trials). We then extracted the core elements of each inclusion/exclusion criterion and summarized these 1,674 eligibility criteria into 678 unique criterion patterns. Many of the inclusion and exclusion criterion patterns were fundamentally similar (i.e., querying the same core data elements). Note that some criteria can be decomposed into multiple sub-criteria; we thus considered the smallest units as individual study traits (e.g., "*history of primary CNS (central nerve system) tumor, or stroke*" can be decomposed into "*history of primary CNS tumor*" and "*history of stroke*"). **Table 3** shows the top 10 most frequent criterion patterns, separated by inclusion vs. exclusion.

**Table 3**. Top 10 frequent criterion patterns used by the 57 trials, separated by inclusion vs. exclusion.

| Rank | Inclusion Criterion Pattern | Study Coverage # of Studies (%) N = 57 | Exclusion Criterion Pattern | Study Coverage # of Studies (%) N = 57 |
|---|---|---|---|---|
| 1 | Aspartate aminotransferase (AST) | 32 (56.14%) | Unstable angina | 36 (63.16%) |
| 2 | Measurable disease | 29 (50.88%) | Myocardial infarction | 34 (59.65%) |
| 3 | Age | 29 (50.88%) | Radiotherapy | 30 (52.63%) |
| 4 | Absolute neutrophil count | 29 (50.88%) | Congestive heart failure | 26 (45.61%) |
| 5 | Platelets | 27 (47.36%) | Pregnant | 22 (38.60%) |
| 6 | Metastatic colorectal cancer | 27 (47.36%) | Bone fracture | 22 (38.60%) |
| 7 | Hemoglobin | 25 (43.86%) | Significant traumatic injury | 22 (38.60%) |
| 8 | Bilirubin | 22 (38.60%) | Bleeding diathesis | 21 (36.84%) |
| 9 | Creatinine | 22 (38.60%) | Chemotherapy | 20 (35.09%) |
| 10 | Alanine transaminase (ALT) | 20 (35.09%) | Skin ulcers | 20 (35.09%) |

However, not all eligibility criteria were computable against our OneFlorida patient database. We found that 194 (28.61%) of the 678 unique patterns were not computable. The main reasons are: (1) the criterion asked for subjective information (e.g., patient's consent or investigator's judgement of patient's health status); and (2) the data elements needed for the criterion were not presented in the OneFlorida data (e.g., "*performance status ecog 0-1*" is not captured).

**The composite patient-trial generalizability (cPTG) score**

Out of the 57 trials, there are 10 Phase I, 6 Phase I/II, 31 Phase II, and 7 Phase III trials (the other 2 have no phase information). We randomly selected 4 trials out of the 7 Phase III trials to model patient eligibility and trial generalizability. We selected Phase III trials as they are conducted to expand on the safety and effectiveness results from Phase I and II trials, to compare the drug to standard therapies, and to evaluate the overall risks and benefits of the treatment, right before the treatment can be approved by the FDA and be put on the market. A Phase III trial typically recruits larger groups of people with more relaxed eligibility criteria comparing to Phase I and II trials.

To calculate cPTG scores, we first calculated the mGIST scores of the 4 trials using population A as the target population (i.e., patients who were treated with Bevacizumab for colorectal cancer). **Table 4** lists the total number of study traits based on the eligibility criteria (regardless of inclusion or exclusion), the number of computable traits, and the mGIST score for each of the 4 trials. There are 219 unique study traits across the 4 trials, and the top 5 common traits are: age, aspartate aminotransferase (AST), bone fracture, pregnancy test, and skin ulcers.

**Table 4**. Characteristics of study traits, mGIST scores, and the relationships between eligibility and number of SAEs.

|  | Total # of traits | # of computable traits | mGIST | Mean # of SAEs | | Wilcoxon Rank Sums Test P value |
|---|---|---|---|---|---|---|
|  |  |  |  | Eligible (SD) | Not eligible (SD) |  |
| NCT00025337 | 46 | 38 | 0.547 | 0.8 (3.0) | 1.5 (5.8) | <.0001 |
| NCT00252564 | 84 | 64 | 0.750 | 1.0 (4.1) | 1.4 (4.5) | 0.0163 |
| NCT00096278 | 88 | 66 | 0.584 | 0.9 (4.1) | 1.4 (4.4) | 0.0013 |
| NCT00112918 | 82 | 57 | 0.307 | 0.5 (2.0) | 1.2 (4.6) | <.0001 |

*SAE: serious adverse event; SD: standard deviation; mGIST: multi- trait Generalizability Index on Study Traits (GIST 2.0).

We first used Wilcoxon rank sums test to test the difference of the number of SAEs between eligible vs. no eligible patients and found the differences are statically significant for all 4 trials (**Table 4**).

**The relationship between the patient-trial generalizability and clinical outcomes**

The zero-inflated negative binomial model results have two parts, as shown in **Table 5**.

**Table 5**. Zero-inflated negative binomial model for the relationships between cPTG and clinical outcomes.

| *Part 1: logistic part for excessive zero (i.e., having no SAE)* | | | |
|---|---|---|---|
| **Parameter** | **Estimate** | **Wald 95% Confidence Interval** | **P value** |
| Age at last Bevacizumab PX | 0.03 | (-0.01, 0.06) | 0.1153 |
| Female vs Male | -0.78 | (-1.4, -0.10) | 0.0242 < 0.05 |
| Race/Ethnicity | | | |
|    Hispanic vs NHW | -0.55 | (-1.37, 0.27) | 0.1882 |
|    NHB vs NHW | -0.35 | (-1.12, 0.42) | 0.3767 |
|    Other vs NHW | 0.12 | (-4.09, 4.33) | 0.956 |
|    Unknown vs NHW | 0.53 | (-0.22, 1.28) | 0.1638 |
| Follow up day | 0.00 | (-0.01, 0.00) | 0.1882 |
| Number of Bevacizumab PXs | -0.05 | (-0.10, -0.01) | 0.0262 < 0.05 |
| cPTG score | 1.30 | (-0.48, 3.08) | 0.1519 |
| *Part 2: negative binomial part* | | | |
| **Parameter** | **Estimate** | **Wald 95% Confidence Interval** | **P value** |
| Age at last Bevacizumab PX | -0.010 | (-0.028, 0.009) | 0.3038 |
| Female vs Male | -0.225 | (-0.586, 0.137) | 0.223 |
| Race/Ethnicity | | | |
|    Hispanic vs NHW | -0.262 | (-0.713, 0.189) | 0.2542 |
|    NHB vs NHW | -0.078 | (-0.521, 0.364) | 0.729 |
|    Other vs NHW | 0.068 | (-2.834, 2.970) | 0.9633 |
|    Unknown vs NHW | 0.057 | (-0.450, 0.564) | 0.8249 |
| Follow up day | 0.003 | (-0.001, 0.006) | 0.1047 |
| Number of Bevacizumab PXs | 0.024 | (0.012, 0.037) | 0.0002 < 0.05 |
| cPTG score | -1.079 | (-1.996, -0.162) | 0.0211 < 0.05 |
| Dispersion | 5.421 | (3.561, 8.253) |  |

The first part is a logistic model, estimating the probability of being an excessive zero (i.e., having no SAE). Two variables—gender and the number of Bevacizumab procedures—are statistically significant at 0.05 level in this part of the model. The odds ratio for female verse male of being an excessive zero is exp(-0.78) = 0.458. This indicates that female has a higher possibility of having SAEs than male while holding other predictors constant. Further, with the number of Bevacizumab procedures increasing by 1, the odds of having no SAE decreases by 0.049 (i.e., 1- exp(-0.05)=0.049). This indicates that having more Bevacizumab procedures increases the possibility of having SAEs, holding other predictors constant. Although the cPTG score is not statistically significant in the logistic model, the estimate equals to 1.30, meaning the odds of having no SAE increase by exp(1.30*0.1) - 1 = 0.139 for a 0.1 increase in the patient-trial generalizability score, holding other variables constant.

The second part is a negative binomial regression model, estimating the expected number of SAEs. Two variables—the number of Bevacizumab procedures and cPTG score—are statistically significant at 0.05 level. The expected number of SAEs increases by exp(0.024) – 1 = 0.0243 for 1 increases in the number of Bevacizumab procedures, indicating more Bevacizumab procedures would have more SAEs. Further, the expected change in the number of SAEs decreases by 1 – exp(-1.079*0.1) = 0.102 for 0.1 increase in cPTG. This indicates that a higher cPTG would lead to a smaller number of SAEs, while holding other variables constant. The dispersion is 5.421 (>1); thus, using a binomial model is more appropriate than a Poisson model.[26]

**Discussion and conclusion**

Our results successfully connect the generalizability to clinical outcomes using RWD. First, we showed that the patients who are eligible for the original trials (that were used to develop the treatment, population B in our case) have better treatment outcomes compared with those who are not eligible (i.e., population C) in real-world clinical practice. This indicates reduced generalizability of the original trials because of the differences between the characteristics of the trial (or trial-eligible) population and ineligible population. Because of these differences, the original trials may have failed to gather sufficient safety and efficacy data on the entire target population, resulting in poorer outcomes when the treatment is applied in real-world clinical settings. Second, we devised a new patient-trial generalizability score considering both patient-level eligibility and trial-level *a priori* generalizability. A higher cPTG score of a patient indicates that the patient can be eligible for more (original) trials (i.e., the patient's characteristics are more similar to those in the trial population) and these trails have a higher *a priori* generalizability.

These results have multiple implications. First, it shows the feasibility of using RWD, especially EHRs, to assess patients' clinical outcomes (e.g., adverse events) and examine various patient characteristics associated with these outcomes. Being able to capture AEs using data collected through routine the standard of care provides us the opportunity to create informatics surveillance systems for post-market drug safety monitoring. Comparing to the existing AE reporting systems (e.g., the FDA Adverse Event Reporting System, [FAERS]), EHRs is a much richer data source (e.g., lab results and diagnoses of other diseases) that provides a more complete picture of patients' characteristics. Further, being able to work with data from a large data research network not only gives us a large sample size and sufficient statistical power to make robust conclusions but also makes our approaches adaptable to other similar networks. Because of the use of a common data model (CDM), our study can be readily replicated in other sites of the PCORnet, which covers more than 100 million patients across the United State.[28] Moreover, it is also possible to extend our study to other clinical research network initiatives such as the National Center for Advancing Translational Sciences (NCATS)'s Clinical and Translational Science Awards Accrual to Clinical Trials (CTSA ACT) network and the Observational Health Data Sciences and Informatics (OHDSI) consortium. Even though they are using different CDMs, adapting our approach to a new CDM is rather straightforward.

Further, our study builds a body of evidence to support the development of an eligibility criteria design tool for optimizing study generalizability at the study design phase. Such an eligibility criteria design tool is much needed and will benefit stakeholders of the clinical trial communities. To the extent that trial participants share the same characteristics as the majority of patients, trails should be developed to fit patient profiles seen in real-world clinical practice, and thus, facilitate the application of trial results to real-world clinical settings.[29] This could be achieved if trials had few restrictions on eligibility, allowing more representative patients to participate. A significant proportion of real world patients are unable to participate in clinical trials due to stringent exclusion criteria, but many still received treatment outside of clinical trials and benefited from therapy.[30] A recent special issue on eligibility criteria in the Journal of Clinical Oncology—a leading clinical oncology research journal—calls for "*broadening eligibility criteria to make clinical trials more representative.*"[31] As a response, the National Cancer Institute (NCI) has revised its clinical trial protocols to expand access for previously excluded patients in an effort to ensure that study participants are more reflective of real-world populations.[32] Our study provides initial data evidence that 1) *a priori* generalizability is qualifiable and 2) combined with patient eligibility, the cPTG score can predict clinical outcomes in real-world patients. This eventually could lead to metrics that rationalize the design (or relaxation) of eligibility criteria.

Last, our study also leads to opportunities to develop a computable eligibility criteria framework for EHR-based cohort identification to facilitate trial recruitment. One can imagine a tool that a trial investigator could not only identify patients who are eligible to the trial of interest based on their EHRs, but also be able to estimate the impact of recruiting a particular patient on the trial's generalizability and possible clinical outcomes on similar patients.

Our study is not without limitations. The process of analyzing and decomposing existing eligibility criteria into computable eligibility criteria against EHRs is both time- and labor-intensive. Exploring advanced natural language processing tools tailored for analyzing eligibility criteria might be beneficial. Nevertheless, to tease out the subtle ambiguities in free-text eligibility criteria, human judgments are always needed. Moreover, many eligibility criteria cannot be accurately translated into database queries (e.g., "*At least 10 days since prior aspirin dose of more than 325 mg/day*"). We took a simplistic approach and did not consider these temporal constraints, which may lead to a small number of inaccurate identifications of patient eligibility. Further, some criteria (e.g., lab results) may be matched to multiple observations (e.g., blood pressure) that vary from time to time. We made an assumption that the patient will meet the criterion as long as if any one of the observations fell into the permissible value range.

In sum, our results are significant. Our ultimate goal is to provide an easy-to-use and more efficient computable eligibility criteria construction platform for investigators to identify eligible patients based on their existing EHRs while maximizing the trial's *a priori* generalizability.

**Acknowledgements**

This work was supported in part by NIH grants AG061431 and UL1TR001427, the OneFlorida Cancer Control Alliance (funded by James and Esther King Biomedical Research Program, Florida Department of Health Grant Number 4KB16), the OneFlorida Clinical Research Consortium (CDRN-1501-26692) funded by the Patient Centered Outcomes Research Institute (PCORI). The content is solely the responsibility of the authors and does not necessarily represent the official views of the NIH or PCORI.

**References**

1. From the NIH Director: The Importance of Clinical Trials | NIH MedlinePlus the Magazine. https://medlineplus.gov/magazine/issues/summer11/articles/summer11pg2-3.html. Accessed March 4, 2019.
2. Rothwell PM. External validity of randomised controlled trials: "to whom do the results of this trial apply?" *Lancet Lond Engl*. 2005;365(9453):82-93. doi:10.1016/S0140-6736(04)17670-8
3. Sedgwick P. External and internal validity in clinical trials. *BMJ*. 2012;344(feb16 1):e1004-e1004. doi:10.1136/bmj.e1004
4. Schoenmaker N, Van Gool WA. The age gap between patients in clinical studies and in the general population: a pitfall for dementia research. *Lancet Neurol*. 2004;3(10):627-630. doi:10.1016/S1474-4422(04)00884-1
5. Sardar MR, Badri M, Prince CT, Seltzer J, Kowey PR. Underrepresentation of women, elderly patients, and racial minorities in the randomized trials used for cardiovascular guidelines. *JAMA Intern Med*. 2014;174(11):1868-1870. doi:10.1001/jamainternmed.2014.4758
6. Battisti NML, Sehovic M, Extermann M. Assessment of the External Validity of the National Comprehensive Cancer Network and European Society for Medical Oncology Guidelines for Non-Small-Cell Lung Cancer in a Population of Patients Aged 80 Years and Older. *Clin Lung Cancer*. 2017;18(5):460-471. doi:10.1016/j.cllc.2017.03.005
7. Beers E, Moerkerken DC, Leufkens HGM, Egberts TCG, Jansen PAF. Participation of older people in preauthorization trials of recently approved medicines. *J Am Geriatr Soc*. 2014;62(10):1883-1890. doi:10.1111/jgs.13067
8. Bellera C, Praud D, Petit-Monéger A, McKelvie-Sebileau P, Soubeyran P, Mathoulin-Pélissier S. Barriers to inclusion of older adults in randomised controlled clinical trials on Non-Hodgkin's lymphoma: a systematic review. *Cancer Treat Rev*. 2013;39(7):812-817. doi:10.1016/j.ctrv.2013.01.007
9. Golomb BA, Chan VT, Evans MA, Koperski S, White HL, Criqui MH. The older the better: are elderly study participants more non-representative? A cross-sectional analysis of clinical trial and observational study samples. *BMJ Open*. 2012;2(6). doi:10.1136/bmjopen-2012-000833
10. Lewis JH, Kilgore ML, Goldman DP, et al. Participation of patients 65 years of age or older in cancer clinical trials. *J Clin Oncol Off J Am Soc Clin Oncol*. 2003;21(7):1383-1389. doi:10.1200/JCO.2003.08.010

11. Wysowski DK, Swartz L. Adverse drug event surveillance and drug withdrawals in the United States, 1969-2002: the importance of reporting suspected reactions. *Arch Intern Med*. 2005;165(12):1363-1369. doi:10.1001/archinte.165.12.1363
12. Bonell C, Oakley A, Hargreaves J, Strange V, Rees R. Assessment of generalisability in trials of health interventions: suggested framework and systematic review. *BMJ*. 2006;333(7563):346-349. doi:10.1136/bmj.333.7563.346
13. Zimmerman M, Chelminski I, Posternak MA. Exclusion criteria used in antidepressant efficacy trials: consistency across studies and representativeness of samples included. *J Nerv Ment Dis*. 2004;192(2):87-94. doi:10.1097/01.nmd.0000110279.23893.82
14. Sen A, Ryan PB, Goldstein A, et al. Correlating eligibility criteria generalizability and adverse events using Big Data for patients and clinical trials. *Ann N Y Acad Sci*. 2017;1387(1):34-43. doi:10.1111/nyas.13195
15. Weng C, Li Y, Ryan P, et al. A Distribution-based Method for Assessing The Differences between Clinical Trial Target Populations and Patient Populations in Electronic Health Records. *Appl Clin Inform*. 2014;5(2):463-479. doi:10.4338/ACI-2013-12-RA-0105
16. He Z, Chandar P, Ryan P, Weng C. Simulation-based Evaluation of the Generalizability Index for Study Traits. *AMIA Annu Symp Proc AMIA Symp*. 2015;2015:594-603.
17. He Z, Ryan P, Hoxha J, et al. Multivariate analysis of the population representativeness of related clinical studies. *J Biomed Inform*. 2016;60:66-76. doi:10.1016/j.jbi.2016.01.007
18. Sen A, Chakrabarti S, Goldstein A, Wang S, Ryan PB, Weng C. GIST 2.0: A scalable multi-trait metric for quantifying population representativeness of individual clinical studies. *J Biomed Inform*. 2016;63:325-336. doi:10.1016/j.jbi.2016.09.003
19. U.S. Food & Drug Administration. *FRAMEWORK FOR FDA'S REAL-WORLD EVIDENCE PROGRAM*.; 2018:40. https://www.fda.gov/downloads/ScienceResearch/SpecialTopics/RealWorldEvidence/UCM627769.pdf. Accessed March 8, 2019.
20. Sherman RE, Anderson SA, Dal Pan GJ, et al. Real-World Evidence — What Is It and What Can It Tell Us? *N Engl J Med*. 2016;375(23):2293-2297. doi:10.1056/NEJMsb1609216
21. Cohen MH, Gootenberg J, Keegan P, Pazdur R. FDA drug approval summary: bevacizumab plus FOLFOX4 as second-line treatment of colorectal cancer. *The Oncologist*. 2007;12(3):356-361. doi:10.1634/theoncologist.12-3-356
22. Shenkman E, Hurt M, Hogan W, et al. OneFlorida Clinical Research Consortium: Linking a Clinical and Translational Science Institute With a Community-Based Distributive Medical Education Model. *Acad Med J Assoc Am Med Coll*. 2018;93(3):451-455. doi:10.1097/ACM.0000000000002029
23. Elizabeth Shenkman, William Hogan. OneFlorida Clinical Research Consortium. https://www.pcori.org/research-results/2015/oneflorida-clinical-research-consortium. Published January 25, 2019. Accessed March 8, 2019.
24. OneFlorida CRC. OneFlorida Deduper: Tools for EHR patient de-duplication (aka entity resolution). https://github.com/ufbmi/onefl-deduper. Published 2018. Accessed March 8, 2019.
25. Jiang Bian, Andrei Sura, Gloria P. Lipori, et al. Implementing a Hash-based Privacy-Preserving Entity Resolution Tool in the OneFlorida Clinical Data Research Network. In: ; 2017.
26. Cameron AC, Trivedi PK. *Regression Analysis of Count Data*. Cambridge, UK ; New York, NY, USA: Cambridge University Press; 1998.
27. Zhang H, He Z, He X, et al. Computable Eligibility Criteria through Ontology-driven Data Access: A Case Study of Hepatitis C Virus Trials. *AMIA Annu Symp Proc AMIA Symp*. 2018;2018:1601-1610.
28. PCORnet. PCORnet Data. https://pcornet.org/pcornet-data/. Published May 1, 2018. Accessed March 12, 2019.
29. Gotay CC. Increasing Trial Generalizability. *J Clin Oncol*. 2006;24(6):846-847. doi:10.1200/JCO.2005.04.5120
30. Karim S, Xu Y, Quan ML, Dort JC, Bouchard-Fortier A, Cheung WY. Generalizability of common cancer clinical trial eligibility criteria in the real world. *J Clin Oncol*. 2018;36(15_suppl):e18616-e18616. doi:10.1200/JCO.2018.36.15_suppl.e18616
31. Kim ES, Bruinooge SS, Roberts S, et al. Broadening Eligibility Criteria to Make Clinical Trials More Representative: American Society of Clinical Oncology and Friends of Cancer Research Joint Research Statement. *J Clin Oncol*. 2017;35(33):3737-3744. doi:10.1200/JCO.2017.73.7916
32. NCI. Inclusion/Exclusion Criteria for National Cancer Institute (NCI) Sponsored Clinical Trials. https://ctep.cancer.gov/protocolDevelopment/docs/NCI_ASCO_Friends_Eligibility_Criteria.pdf. Published September 26, 2018. Accessed March 12, 2019.